\documentclass[prd,twocolumn,showpacs,floats,floatfix,nofootinbib]{revtex4}
\usepackage{graphicx}
\usepackage{dcolumn}
\usepackage{amssymb}
\usepackage{bm}
\def\spose#1{\hbox to 0pt{#1\hss}}

\def\lta{\mathrel{\spose{\lower 3pt\hbox{$\mathchar"218$}}
     \raise 2.0pt\hbox{$\mathchar"13C$}}}
\def\gta{\mathrel{\spose{\lower 3pt\hbox{$\mathchar"218$}}
     \raise 2.0pt\hbox{$\mathchar"13E$}}}
\newcommand{\be}{\begin{equation}}
\newcommand{\en}{\end{equation}}
\newcommand{\bea}{\begin{eqnarray}}
\newcommand{\ena}{\end{eqnarray}}
\newcommand{\ex}{\mbox{e}}
\newcommand{\dd}{\mbox{d}}

\def\setR{\mathbb{R}}
\def\setC{\mathbb{C}}

\newcommand{\ie}{\textsl{i.e.~}}

\newcommand{\eg}{\textsl{e.g.~}}
\newcommand{\etal}{\textsl{et al.~}}

 \newcommand{\Ka}{{\cal K}}
 
\newcommand{\GN}{G_{_\mathrm{N}}}

\newcommand{\lP}{\ell_{_\mathrm{Pl}}}

\begin{document}

\title{Gravitational wave background in perfect fluid quantum
cosmologies}

\author{Patrick Peter} \email{peter@iap.fr}
\affiliation{${\cal G}\setR\varepsilon\setC{\cal O}$ --
Institut d'Astrophysique de
 Paris, UMR7095 CNRS, Universit\'e Pierre \& Marie Curie, 98 bis
 boulevard Arago, 75014 Paris, France}

\author{Emanuel J. C. Pinho} \email{emanuel@cbpf.br}
\affiliation{Lafex - Centro Brasileiro de Pesquisas F\'{\i}sicas --
CBPF, \\ rua Xavier Sigaud, 150, Urca, CEP22290-180, Rio de Janeiro,
Brazil}

\author{Nelson Pinto-Neto} \email{nelsonpn@cbpf.br} \affiliation{Lafex
- Centro Brasileiro de Pesquisas F\'{\i}sicas -- CBPF, \\ rua Xavier
Sigaud, 150, Urca, CEP22290-180, Rio de Janeiro, Brazil}

\date{\today}

\begin{abstract}
We discuss the gravitational wave background produced by bouncing
models based on a full quantum evolution of a universe filled with a
perfect fluid. Using an ontological interpretation for the background
wave function allows us to solve the mode equations for the tensorial
perturbations, and we find the spectral index as a function of the
fluid equation of state.
\end{abstract}

\maketitle

\section{Introduction}

The theory of cosmological perturbations~\cite{MFB} relies essentially
on two assumptions, namely that the background is described by pure
classical General Relativity (GR), while the perturbations thereof
stem from quantum fluctuations, although they are subsequently evolved
classically. Quite apart from the computational usefulness of this
scheme, this state of affairs is rather incomplete, and one would
expect instead a fully quantum treatment of both the background and
the perturbations to be achievable. In fact, the overwhelming majority
of classical backgrounds possess an initial singularity at which the
classical theory is expected to break down. In recent years, many
quantum background cosmological models have been proposed, which share
the attractive property of exhibiting neither singularities nor
horizons~\cite{pinto,pinto2,fabris}, leading the universe evolution
through a bouncing phase due to quantum effects, and a contracting
phase from infinity before the bounce. These new features of the
background introduce a new picture for the evolution of cosmological
perturbations: vacuum initial conditions may now be imposed when the
Universe was very big and almost flat, and effects due to the
contracting and bouncing phases, which are not present in the standard
background model, may change the subsequent evolution of perturbations
in the expanding phase.  Hence, it is quite important to study the
evolution of perturbations in these quantum backgrounds.  The aim of
the present paper is to provide a step in this direction by
considering tensor perturbations in quantum minisuperspace background
solutions.  Interpreting the quantum theory in an ontological
way~\cite{onto,santini} allows to define quantum scale factor
trajectories, which can then be used in the second order tensorial
modes perturbation equations as show in Ref.~\cite{tens1}.

Note that such models may be viewed as alternatives to the standard
inflationary paradigm~\cite{inflation}.  Most known
alternatives~\cite{PBB,ekp} to inflation present a primordial bouncing
phase~\cite{bounce,jmpp}. Note that such a phase can also be seen as a
complementary ingredient necessary for a complete cosmological
scenario to make actual sense, \ie not to be plagued with a
singularity~\cite{singularity}, or to avoid facing any trans-Planckian
problem~\cite{transP} if, for instance, the bounce occurs at a scale such
that all relevant cosmological scales now never went trans-Planckian.
The bounce phase has recently been the subject
of a lot of attention, in particular in view of the fact that in many
instances, it was found to have the ability to modify the primordial
spectrum of scalar perturbations, thus paving the way to confront them
to the observational data~\cite{WMAP}.  In the case of bounces in
quantum cosmological models, although the evolution equations for the
perturbations may be constructed~\cite{halli}, they are rather
complicated due to the fact that the background does not satisfy
classical Einstein's equations. Hence, all works in this area had to
rely on a semiclassical approximation.

In this paper, we calculate the gravitational wave background spectrum
produced at the bounce transition when this phase is described by a
perfect fluid and the theory is fully quantized: this is the first
time such a calculation, not involving any semiclassical approximation
is performed. The restriction to gravitational waves stems from the
fact that the perturbation equations for this type of modes can be
substantially simplified, even when the background is
quantized~\cite{tens1}. Scalar and vector modes, however, exhibit
technical difficulties which have not been solved yet, so that tensor
modes are, for the time being, the only modes that can be studied in a
completely quantum way.

The paper is organized as follows. Recalling in Sec.~\ref{sec:mode}
how the ontological interpretation allows a simple separation between
the background and the perturbations, we explain in
Sec.~\ref{sec:traj} how the Bohmian trajectories for the scale factor
are derived, and discuss their generality. Then, Sec.~\ref{sec:mu},
which is the core of this work, provides the tensorial modes indices
in the known cases. We end in Sec.~V with conclusions and discussions.

\section{The mode equation}\label{sec:mode}

The action we shall begin with is that of GR with a perfect fluid, the
latter being described by the formalism due to Schutz~\cite{Schutz},
\ie
\begin{equation}
\mathcal{S}= \mathcal{S}_{_\mathrm{GR}} + \mathcal{S}_\mathrm{fluid} =
-\frac{1}{6\lP^2} \int \sqrt{-g} R \dd^4 x + \int \sqrt{-g} p\dd^4 x,
\label{action}
\end{equation}
where $\lP=(8\pi\GN/3)^{1/2}$ is the Planck length in natural units
($\hbar=c=1$), $p$ is the perfect fluid pressure whose density $\rho$
is provided by the relation $p=\omega\rho$, $\omega$ being a
constant. The metric $\mbox{\boldmath$g$\unboldmath}$ in
Eq.~(\ref{action}) is of the Friedman-Lema\^{\i}tre-Robertson-Walker
(FLRW) type, whose line element we choose to be given by
\begin{equation}
\dd s^{2}=N^2\left(\tau\right)\dd\tau^2 -a_{\rm
  phys}^2\left(\tau\right) \left(\gamma_{ij}+w_{ij}\right)\dd x^{i}\dd
x^{j},
\label{adm2}
\end{equation}
\ie we assume it is perturbed to first order and restrict attention to
tensorial perturbations only, with $w^{ij}_{\ \ |j}=0$ and $w^i_{\
i}=0$, indices being raised and lowered by means of the background
metric $\gamma_{ij}$ of the spacelike hypersurfaces (the bar denotes a
covariant derivative with respect to this metric); the lapse function
$N(\tau)$, once fixed, defines the gauge.

After inserting Eq.~(\ref{adm2}) into the action~(\ref{action}), and
performing Legendre and canonical transformations, the Hamiltonian up
to second order reads (see Ref.~\cite{tens1} for details)
\begin{widetext}
\begin{eqnarray}H&\equiv&NH_0\nonumber \\
&=&N\left\{-\frac{P_a^{2}}{4a}-\Ka a+ \frac{P_{_T}}{a^{3\omega}}
\left(1+\displaystyle\frac{\omega}{4} \int \dd^{3} x \gamma^{1/2}\,
w_{ij} w^{ij}\right) + \frac{5P_a^2}{48a} \int \dd^3 x \gamma^{1/2}\,
w_{ij}w^{ij} \right. \nonumber\\&& + \left. \int \dd^3 x \left[
\frac{6\Pi_{ij} {{\Pi}}\null^{ij}}{a^{3} \gamma^{1/2}} + 2\frac{P_a
w_{ij}{\Pi}^{ij}}{a^2} + \gamma^{1/2}a \left( \frac{w^{ij|k} w_{ij|k}
}{24} + \frac{\Ka}{6} w_{ij} w^{ij} \right)\right] \right\},
\label{h1}
\end{eqnarray}
which is nothing but the Hamiltonian of Ref.~\cite{halli} expressed
for a perfect fluid. In Eq.~(\ref{h1}) and in what follows, we shall
denote by $\Ka$ the spatial curvature ($\Ka=0,\pm 1$ for flat, open
and closed space respectively) in order to avoid confusion with the
wavenumber $k$ below. The quantities $P_a, {\Pi}^{ij}, P_{_T}$ are the
momenta canonically conjugate to the scale factor, the tensor
perturbations, and to the fluid degree of freedom, respectively.
These quantities have been redefined in order to be dimensionless.
For instance, the physical scale factor $a_{\rm phys}$ can be obtained
from the adimensional $a$ present in~(\ref{h1}) through $a_{\rm
phys}=\lP a/\sqrt{V}$, where $V$ is the comoving volume of the
background spacelike hypersurfaces.  This Hamiltonian, which is zero
due to the constraint $H_0\approx0$, yields the correct Einstein
equations both at zeroth and first order in the perturbations, as can
be checked explicitly. In order to obtain its expression, no
assumption has been made about the background dynamics.

In the quantum regime, this Hamiltonian can be substantially
simplified through the implementation of the quantum canonical
transformation generated by
\begin{equation}
\label{u2}
U=\exp(iG_{\rm q})
\equiv\exp\left(\frac{i}{12}\hat{\beta}_a \hat{Q}\right),
\end{equation}
where $\hat{\beta}_a \equiv \frac{1}{2} \left(\hat{P}_a \hat{a} +
\hat{a}\hat{P}_a\right)$ and $\hat{Q} \equiv \int \dd^3 x\,
\gamma^{1/2} \hat{w}_{ij} \hat{w}^{ij}$ are the self-adjoint operators
associated with the corresponding classical variables, yielding, for a
particular factor ordering of~(\ref{h1}) (see Ref.~\cite{tens1} for
details),
\begin{equation}
\hat{H}_0 = \left[ -\frac{1}{4\hat{a}}\hat{P}_a^2 -\Ka\hat{a}+
\frac{\hat{P}_{_T}}{\hat{a}^{3\omega}} + \int \dd^3 x \left(
6\frac{\hat\Pi^{ij}\hat{\Pi}_{ij}}{\gamma^{1/2} a^3} +
\frac{1}{24} \gamma^{1/2} a \hat{w}_{ij|k}\hat{w}^{ij|k}+
\frac{1}{12} \gamma^{1/2} \Ka \hat{w}_{ij} \hat{w}^{ij} a
\right)\right].
\label{h200}
\end{equation}

As we are here also quantizing the background, the quantization
procedure is now to impose ${\hat{H}}_0\Psi(a,w_{ij})=0$.
The Wheeler-DeWitt equation in this
case reads

\begin{eqnarray}
i\frac{\partial\Psi}{\partial T} & = & \hat{H}_\mathrm{red} \Psi
\nonumber \\
&:=&\left\{ \frac{a^{3\omega-1}}{4} \frac{\partial^2}{\partial a^{2}}
- \Ka a^{3\omega+1}+ \int \dd^3 x \left[ - 6
\frac{a^{3(\omega-1)}}{\gamma^{1/2}} \frac{\delta^2}{\delta w_{ij}
\delta w^{ij}} + a^{3\omega+1} \left(\gamma^{1/2} \frac{w_{ij|k}
w^{ij|k}}{24} + \Ka\frac{w_{ij} w^{ij}}{12}\right)\right]\right\}\Psi,
\label{es2}
\end{eqnarray}
where we have chosen $T$ as the time variable, which is equivalent to
impose the time gauge $N=a^{3\omega}$. Note that such a choice is
possible in the case at hand because we are considering a a perfect
fluid, for which one can use the variable which describes the fluid as
a clock~\cite{time,kuchar}.

Now, if one uses an ontological interpretation of
quantum mechanics like the one suggested by de Broglie and
Bohm~\cite{onto}, and makes the separation ansatz for the wave
functional $\Psi[a,w_{ij},T]=\varphi(a,T)\psi[a,w_{ij},T]$, with
$\psi[a,w_{ij},T]=\psi_1[w_{ij},T]\int \dd a \varphi^{-2}(a,T)
+\psi_2[w_{ij},T]$, then Eq.~(\ref{es2}) can be split into two,
yielding
\begin{equation}
i\frac{\partial\varphi}{\partial T}
=\frac{a^{3\omega-1}}{4} \frac{\partial^2\varphi}{\partial a^2}-
\Ka a^{3\omega+1}\varphi,
\label{es20}
\end{equation}
and
\begin{equation}
i\frac{\partial\psi}{\partial T}=
\int \dd^3 x \left[-6\frac{a^{3(\omega-1)}}{\gamma^{1/2}}
\frac{\delta^2}{\delta w_{ij} \delta w^{ij}} + a^{3\omega+1}
\left(\gamma^{1/2} \frac{w_{ij|k} w^{ij|k}}{24} +
\Ka\frac{w_{ij} w^{ij}}{12}\right)\right]\psi.
\label{es22}
\end{equation}

Using the Bohm interpretation, Eq.~(\ref{es20}) can now be solved as
in Ref.~\cite{pinto,pinto2,fabris}, yielding a Bohmian quantum
trajectory $a(T)$, which in turn can be used in
Eq.~(\ref{es22}). Indeed, since one can view $a(T)$ as a function of
$T$, it is possible to implement the canonical transformation
generated by
\begin{equation}
\label{qct0}
U = \exp\left[ i \left( \int \dd^3 x \gamma^{1/2} \frac{\dot{a}
w_{ij} w^{ij}}{2a} \right) \right] \exp\left\{ i \left[ \int \dd^3 x
\left( \frac{w_{ij}\Pi ^{ij} + \Pi ^{ij} w_{ij}}{2} \right) \ln\left(
\frac{\sqrt{12}}{a} \right) \right]\right\},
\end{equation}
where, as $a(T)$ is a given quantum trajectory coming from Eq.~(\ref{es20}),
Eq.~(\ref{qct0}) must be viewed as the generator of a time dependent
canonical transformation to Eq.~(\ref{es22}). It yields
\begin{equation}
i\frac{\partial\psi}{\partial T}=
\int \dd^3 x \left[-\frac{a^{3\omega-1}}{2\gamma^{1/2}}
\frac{\delta^2}{\delta \mu_{ij}\delta \mu^{ij}} + a^{3\omega-1}
\left(\gamma^{1/2} \frac{\mu_{ij|k} \mu^{ij|k}}{2} +
\Ka \mu_{ij}\mu^{ij} - \frac{\ddot{a}}{2a} \mu_{ij}
\mu^{ij}\right)\right] \psi.
\label{es27}
\end{equation}
Through the redefinition of time $a^{3\omega-1}\dd T=\dd\eta$, we
obtain
\begin{equation}
i\frac{\partial\psi(\mu_{ij},\eta)}{\partial \eta}=
\int \dd^3 x \left\{-\frac{1}{2\gamma^{1/2}}
\frac{\delta^2}{\delta \mu_{ij}\delta \mu^{ij}} +
\gamma^{1/2}\left[\frac{1}{2}\mu_{ij|k} \mu^{ij|k} + \left( \Ka-
\frac{\ddot{a}}{2a}\right) \mu_{ij} \mu^{ij}\right] \right\}
\psi(\mu_{ij},\eta).
\label{NENSE}
\end{equation}
This is the most simple form of the Schr\"odinger equation which
governs tensor perturbations for a quantum minisuperspace model
with fluid matter source.

The equation for the modes $\mu_k = \omega_k/a$ which can be derived
from Eq.~(\ref{NENSE}) reads (for that point on, the $k$-index will be
omitted)

\begin{equation}
\label{mu}
\mu''+\left( k^2 +2\Ka -\frac{a''}{a} \right)\mu =0,
\end{equation}
which has the same form as the one obtained for classical backgrounds
(see Ref.~\cite{MFB}), with the important difference that the function
$a(\eta)$ is no longer the classical solution for the scale factor,
but the quantum Bohmian solution. In this way, we can proceed with the
usual analysis, with the quantum Bohmian solution $a(\eta)$ coming
from Eq.~(\ref{es20}) acting as the new pump field.

\section{The background Bohmian trajectories}\label{sec:traj}

In order to obtain the background quantum solutions, we choose the
following factor ordering for the kinetic term of the background
Schr\"odinger equation (from now on $T=t$):
\begin{eqnarray}
i\frac{\partial\varphi}{\partial t}&=&-\frac{1}{4}\left[
a^{(3\omega-1)/2} \hat{P}_a a^{(3\omega-1)/2}\hat{P}_a\right] \varphi
- \Ka a^{3\omega+1}\varphi \nonumber \\ &=&\frac{1}{4}
\left\{ a^{(3\omega-1)/2}\frac{\partial}{\partial a}
\left[ a^{(3\omega-1)/2}\frac{\partial}{\partial
a}\right] \right\}\varphi- \Ka a^{3\omega+1}\varphi,
\label{es201}
\end{eqnarray}
which is the factor ordering yielding a covariant Schr\"odinger
equation under field redefinitions~\cite{fo}.

The quantum Bohmian trajectories are obtained from the solutions
of Eq.~(\ref{es201}).
There are two distinct situations: whether the spacelike hypersurfaces
are flat or not.
\end{widetext}

\subsection{Flat spatial sections}

With the factor ordering chosen in Eq.~(\ref{es201}) with $\Ka =0$, we
can change variables to $\chi=\frac{2}{3} (1-\omega)^{-1}
a^{3(1-\omega)/2}$, obtaining the simple equation
\begin{equation}
i\frac{\partial\varphi}{\partial t}= \frac{1}{4}
\frac{\partial^2\varphi}{\partial \chi^2}.
\label{es202}
\end{equation}
Note that this is just the time reversed Schr\"odinger equation for a
one dimensional free particle constrained to the positive axis.  As
$a$ and $\chi$ are positive, solutions which have unitary evolution
must satisfy the condition
\begin{equation}
\label{cond27}
\varphi^{\star}\frac{\partial\varphi}{\partial \chi}
-\varphi\frac{\partial\varphi^{\star}}{\partial
  \chi}\Biggl|_{\chi=0}=0
\end{equation}
(see Ref.~\cite{fabris} for details).

We will choose the initial normalized wave function
\begin{equation}
\label{initial}
\varphi_0(\chi)=\biggl(\frac{8}{t_0\pi}\biggr)^{1/4}
\exp\left(-\frac{\chi^2}{t_0}\right) ,
\end{equation}
where $t_0$ is an arbitrary constant which determines the width
of the Gaussian and hence the probability amplitude of initial scale factors.
The Gaussian $\varphi_0$ satisfies condition (\ref{cond27}). It is
a commonly used initial condition when the time gauge is fixed and one
gets a Schr\"odinger equation of the type of Eq.~(\ref{es202})
\cite{pinto,fabris,time}, and even when the time gauge is not fixed
when constructing wave packets \cite{pinto2,kiefer}.

Using the propagator procedure of Refs.~\cite{pinto,fabris}, we obtain
the wave solution for all times in terms of $a$:
\begin{widetext}
\begin{equation}\label{psi1t}
\varphi(a,t)=\left[\frac{8 t_0}{\pi\left(t^2+t_0^2\right)}
\right]^{1/4}
\exp\biggl[\frac{-4t_0a^{3(1-\omega)}}{9(t^2+t_0^2)(1-\omega)^2}\biggr]
\exp\left\{-i\left[\frac{4ta^{3(1-\omega)}}{9(t^2+t_0^2)(1-\omega)^2}
+\frac{1}{2}\arctan\biggl(\frac{t_0}{t}\biggr)-\frac{\pi}{4}\right]\right\}
.
\end{equation}
\end{widetext}

Due to the chosen factor ordering, the probability density $\rho(a,t)$
has a non trivial measure and it is given by
$\rho(a,t)=a^{(1-3\omega)/2}\left|\varphi(a,t)\right|^2$.  Its continuity
equation coming from Eq.~(\ref{es201}) reads
\begin{equation}
\label{cont}
\frac{\partial\rho}{\partial t}
-\frac{\partial}{\partial a}\biggl[\frac{a^{(3\omega-1)}}{2}
\frac{\partial S}{\partial a}\rho\biggr]=0 ,
\end{equation}
which implies in the Bohm interpretation that
\begin{equation}
\label{guidance}
\dot{a}=-\frac{a^{(3\omega-1)}}{2}
\frac{\partial S}{\partial a} ,
\end{equation}
in accordance with the classical relations $\dot{a}=\{a,H\}=
-\frac{1}{2}a^{(3\omega-1)}P_a$ and $P_a=\partial S/\partial a$.

Inserting the phase of (\ref{psi1t}) into Eq.~(\ref{guidance}),
we obtain the Bohmian quantum trajectory for the scale factor:
\begin{equation}
\label{at}
a(t) = a_0
\left[1+\left(\frac{t}{t_0}\right)^2\right]^\frac{1}{3(1-\omega)} .
\end{equation}
Note that this solution has no singularities and tends to the
classical solution when $t\rightarrow\pm\infty$. Remember that we are
in the gauge $N=a^{3\omega}$, and $t$ is related to conformal time
through
\begin{equation}
\label{jauge}
N\dd t = a \dd \eta \quad \Longrightarrow \dd\eta =
\left[a(t)\right]^{3\omega-1} \dd t.
\end{equation}
The solution (\ref{at}) can be obtained for other initial wave functions
(see Ref.~\cite{fabris}).

\subsection{Curved spatial sections}

In this case, only for $\omega=\frac{1}{3}$ (radiation) are there
analytic solutions available. Here, $t=\eta$, and there is no factor
ordering ambiguity in the kinetic term. Equation (\ref{es20}) [or
(\ref{es201})] reduces to the time reversed Schr\"odinger equation for
harmonic or anharmonic oscillators.  Now the condition for unitary
evolution reads
\begin{equation}
\varphi^{\star}\frac{\partial\varphi}{\partial a}
-\varphi\frac{\partial\varphi^{\star}}{\partial a}\Bigg|_{a=0}=0,
\end{equation}
the probability density $\rho(a,t)$ is the trivial one, namely
$\rho(a,t)=\left|\varphi(a,t)\right|^2$, satisfying the continuity
equation
\begin{equation}
\label{cont2}
\frac{\partial\rho}{\partial t}
-\frac{\partial}{\partial a}\left(\frac{1}{2}
\frac{\partial S}{\partial a}\rho\right)=0 ,
\end{equation}
yielding the guidance relation
\begin{equation}
\label{guidance2}
\dot{a}=-\frac{1}{2}
\frac{\partial S}{\partial a} .
\end{equation}

Given the same initial wave function as before, we obtain (see
Ref.~\cite{pinto}),
\begin{widetext}
\begin{eqnarray}\label{psikt}
\varphi(a,\eta)&=&\left\{\frac{8 \eta_0\mathcal{K}}
{\pi\left[\eta_0^2\mathcal{K}\cos^2\left(\sqrt{\mathcal{K}}\eta\right)
+ \sin^2\left( \sqrt{\mathcal{K}}\eta\right) \right]}\right\}^{1/4}
\exp\left[\frac{-\eta_0 a^{2}\mathcal{K}}
{\eta_0^2\mathcal{K}\cos^2\left( \sqrt{\mathcal{K}}\eta\right) +
\sin^2\left( \sqrt{\mathcal{K}}\eta\right)}\right]\times\nonumber \\ &
&\times\exp\left(-i\left\{\frac{\left(1-\mathcal{K} \eta_0^2\right)
\sqrt{\mathcal{K}}a^2 \cos \left( \sqrt{\mathcal{K}}\eta\right)
\sin\left( \sqrt{\mathcal{K}}\eta\right)} {\left[
\eta_0^2\mathcal{K}\cos^2\left( \sqrt{\mathcal{K}}\eta\right) +
\sin^2\left( \sqrt{\mathcal{K}}\eta\right)\right]}
+\frac{1}{2}\arctan\left[\frac{\eta_0\sqrt{\mathcal{K}}
\cos\left(\sqrt{\mathcal{K}}\eta\right)} {\sin\left(
\sqrt{\mathcal{K}}\eta\right)}\right]- \frac{\pi}{4}\right\}\right)
\end{eqnarray}
(we change $t_0$ to $\eta_0$ when $\omega=\frac{1}{3}$).

The Bohmian quantum scale factor obtained through the substitution of
the phase of Eq.~(\ref{psikt}) into Eq.~(\ref{guidance2}) reads
\begin{equation}
\label{atk}
a(\eta) = a_0 \left[1+\frac{\left( 1-\mathcal{K}\eta_0^2\right)
\sin^2\left(
\sqrt{\mathcal{K}}\eta\right)}{\mathcal{K}\eta_0^2}\right]^\frac{1}{2}=
a_0 \left[\cos^2\left(\sqrt{\mathcal{K}}\eta\right)+
\frac{\sin^2\left(\sqrt{\mathcal{K}}\eta\right)}{\mathcal{K}\eta_0^2}
\right]^\frac{1}{2}
\end{equation}
\end{widetext}

For $\mathcal{K}=0$ and radiation, we can obtain the wave solution and
Bohmian trajectories either by taking the respective limits from
Eqs.~(\ref{psi1t}) and (\ref{at}) or Eqs~(\ref{psikt}) and (\ref{atk}).
The resulting Bohmian scale factor is
\begin{equation}
\label{atKp}
a=a_0\sqrt{1+\left(\frac{\eta}{\eta_0}\right)^2}.
\end{equation}

Note that, for the curved space section solutions be realistic, there
must have a long epoch after the bounce when the scale factor recover
its classical evolution and the curvature is negligible, i.e. when the
scale factor in Eq.~(\ref{atk}) can be approximated in some large
interval of $\eta$ to $a(\eta)\propto\eta$ in order for the model to
be compatible with standard nucleosynthesis and cosmological
observations.  This can be accomplished if $\eta_0\ll 1$. It means
that the initial wave function (\ref{initial}) must be a very centered
Gaussian around zero. The flatness problem is then translated to the
quantum cosmological language to the question: why an initial Gaussian
wave function of the Universe is so centered around a null value for
the scale factor?

\section{Tensorial modes propagation}\label{sec:mu}

Having obtained in the previous sections the propagation equation for
the full quantum tensorial modes, namely Eq.~(\ref{mu}), in the
Bohmian picture with the scale factor assuming the form (\ref{at}) or
(\ref{atk}), it is our goal in this section to solve this equation in
order to obtain the gravitational wave power spectrum as predicted by
such models.  The first two subsections deal with the flat spatial
section case for $-\frac{1}{3} < \omega < 1$. The final one treat the
curved spatial cases for $\omega=\frac{1}{3}$.

\subsection{Power spectrum for flat spatial section}

\begin{figure*}[t]
\includegraphics[width=8.5cm]{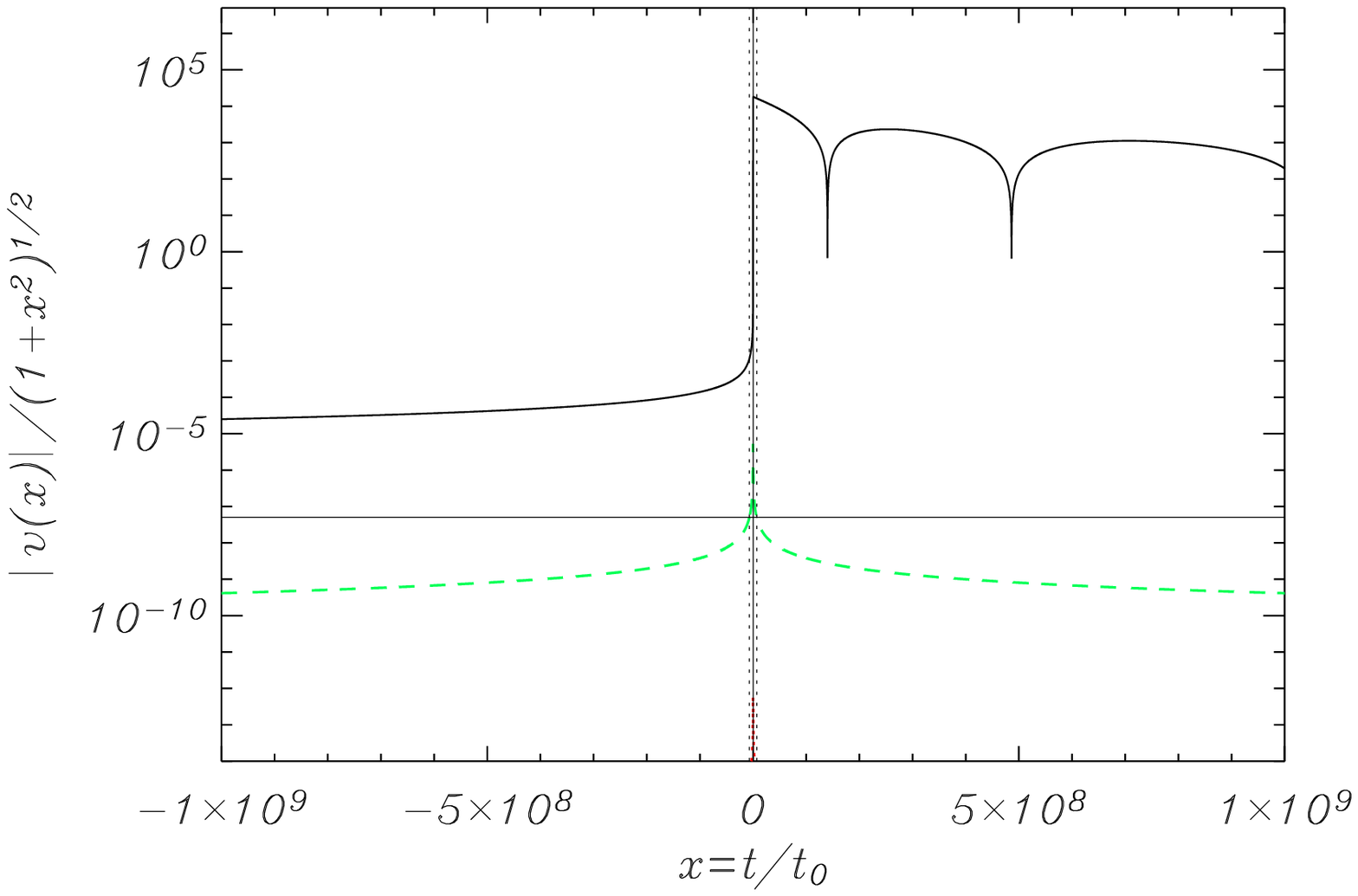}
\includegraphics[width=8.5cm]{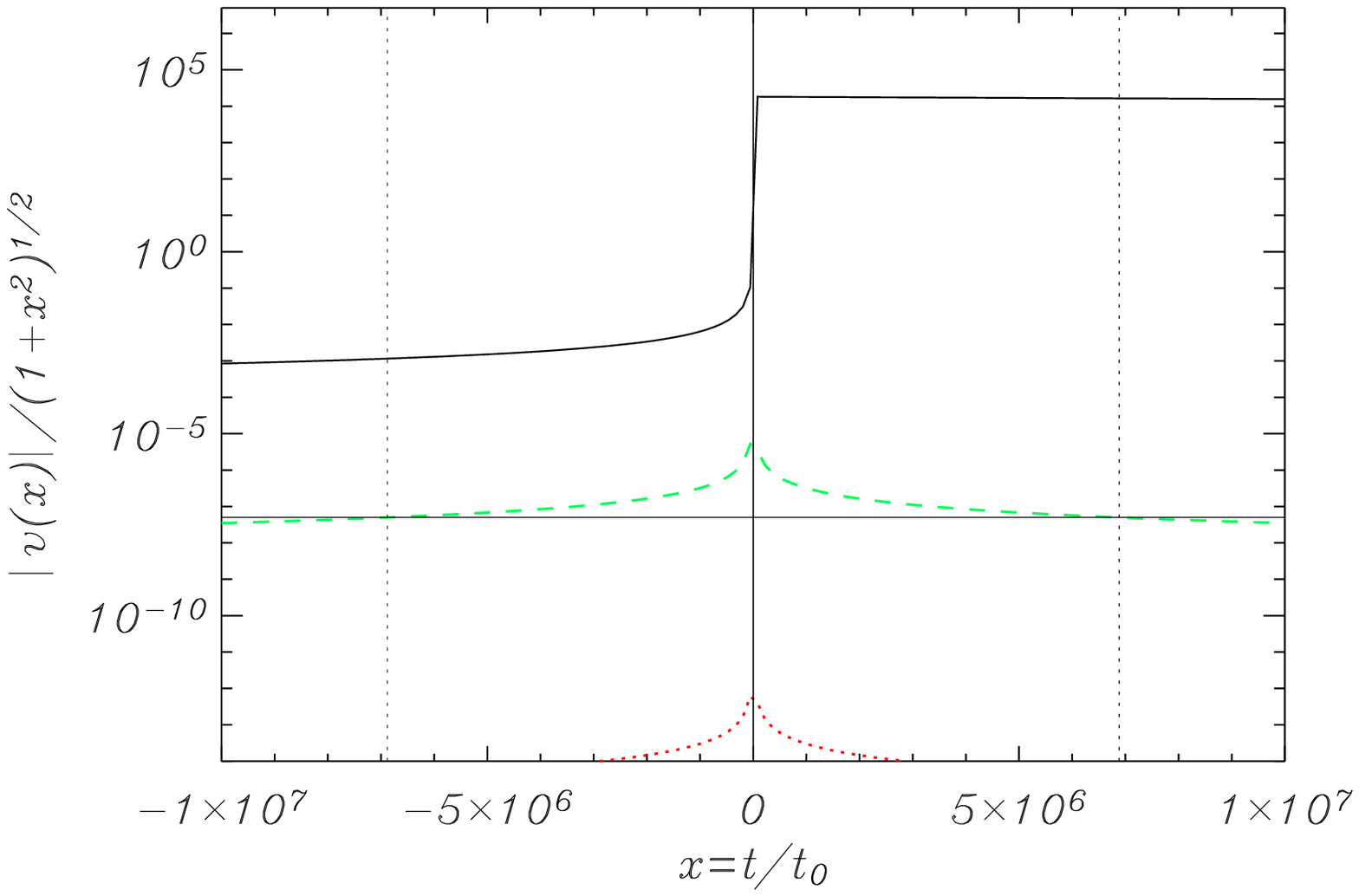}
\caption{Time evolution of the mode function $\bar v$ for the equation
of state $\omega=0.1$ and wavenumber $\tilde k^2=5\times 10^{-8}$. The
full line is $|\bar v(x)|/\sqrt{1+x^2}=\mu/a$ and thus provides
directly the power spectrum. The symmetric curves are background
functions: dashed is the conformal time potential $a''/a$ as given by
Eq.~(\ref{Vmu}), dotted is the term proportional to $\tilde k^2$ in
Eq.~(\ref{v}) and full is $(1+x^2)^{-2}$. The horizontal thin straight
line gives the value of $\tilde k^2$ used to compute the figures. The
left panel shows the full time evolution which was computed. For
$x<0$, there are oscillations only in the real and imaginary parts of
the mode, so the amplitude shown is a non oscillating function of
time. It however acquires an oscillating piece after the bounce has
taken place. The right panel is merely a zoom for smaller time scales
also showing $\pm x_\mathrm{M}$ (the dotted vertical line) and
$x_\mathrm{exit}$ (the full vertical line, indistinguishable on that
scale with the axis). One clearly sees that even though the mode
indeed starts oscillating, it does so on a timescale such that it is
approximately constant all the way to $x_\mathrm{M}$.}
\label{fig:v01}
\end{figure*}

\begin{figure*}[t]
\includegraphics[width=8.5cm]{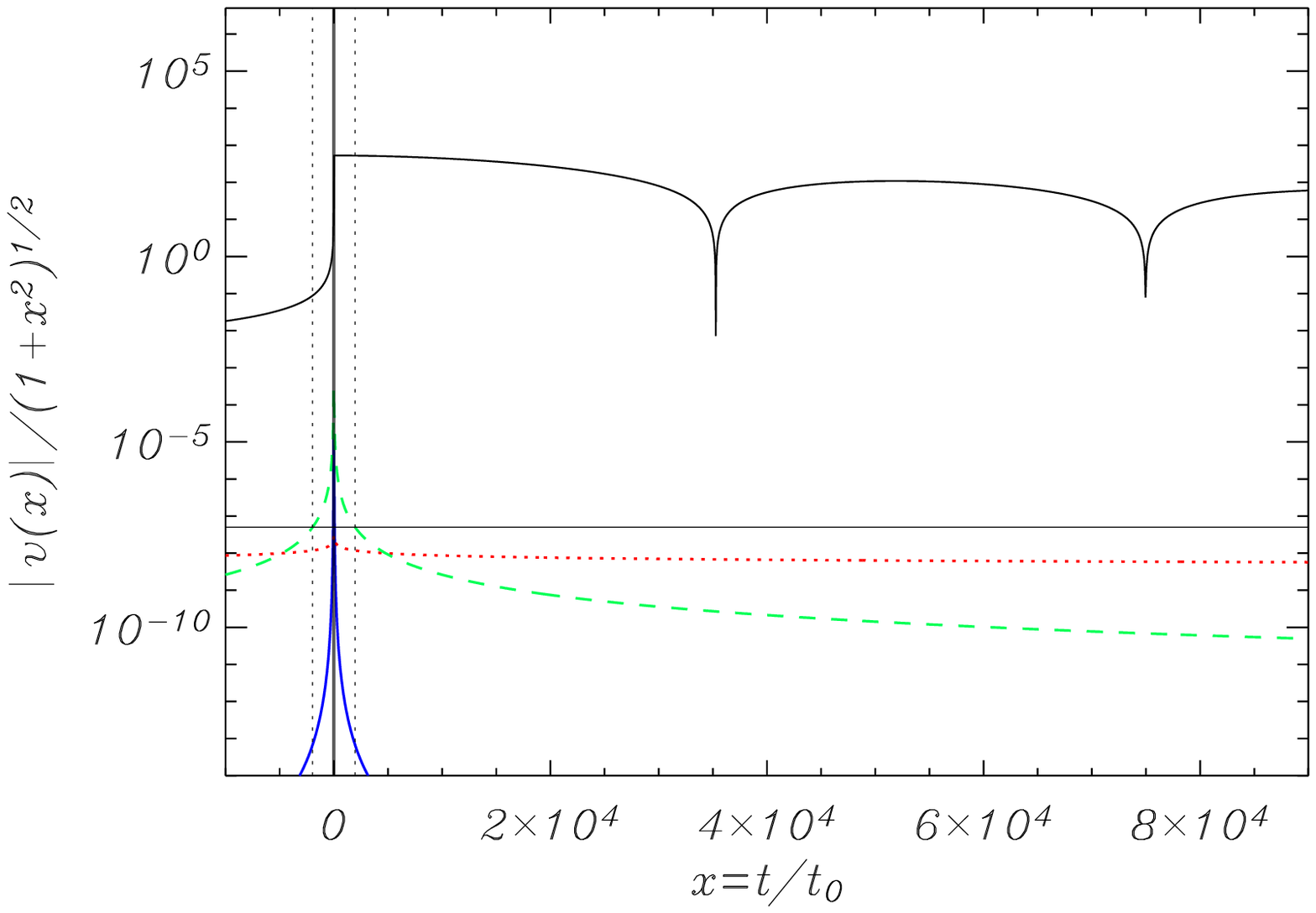}
\includegraphics[width=8.5cm]{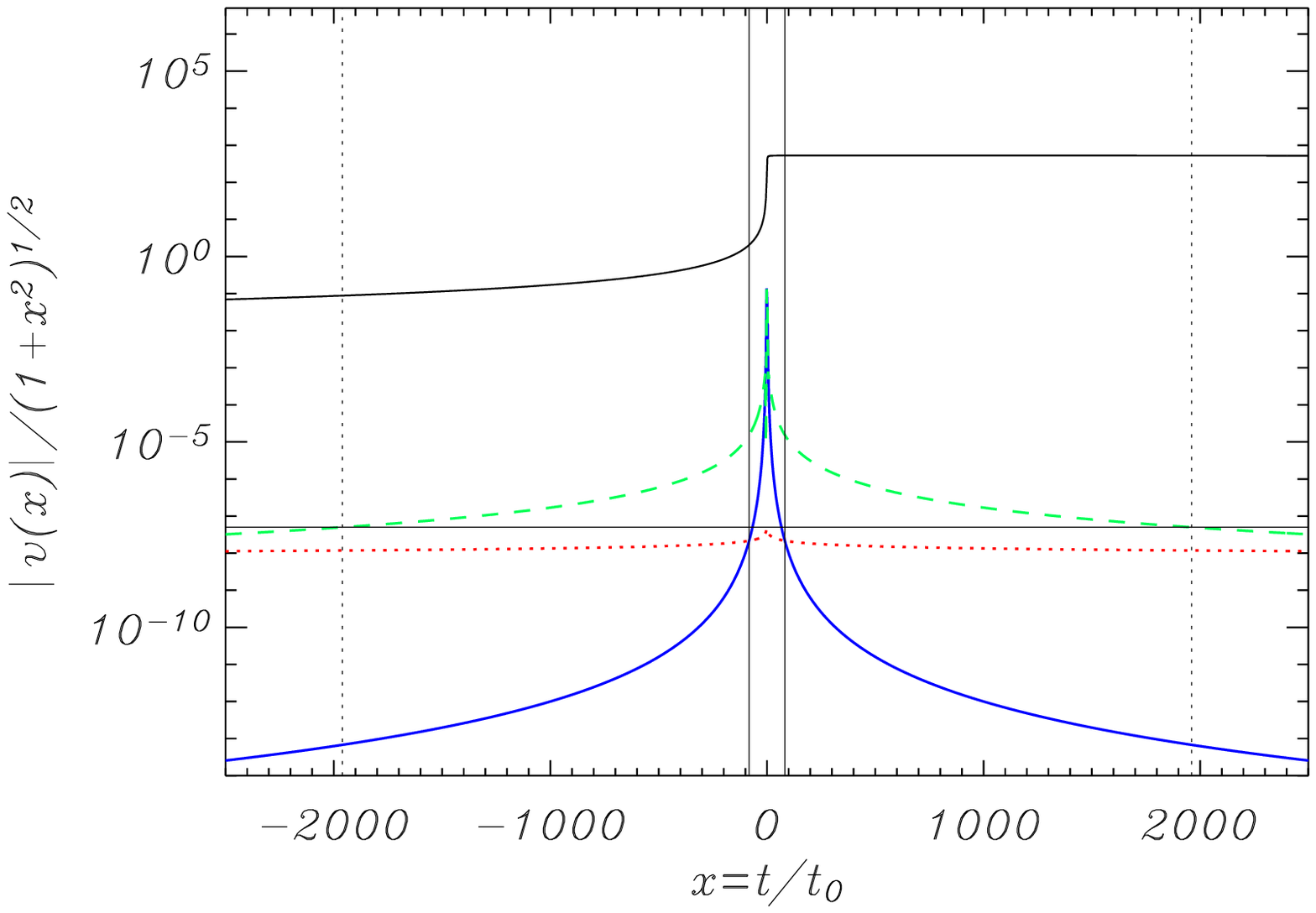}
\caption{Same as Fig.~\ref{fig:v01} with $\omega=0.3$. The
oscillations are visible on the left panel and the detailed view of
the right panel emphasizes that there is no discontinuity on the
mode. It also shows both matching points.}
\label{fig:v03}
\end{figure*}

\begin{figure*}[t]
\includegraphics[width=8.5cm]{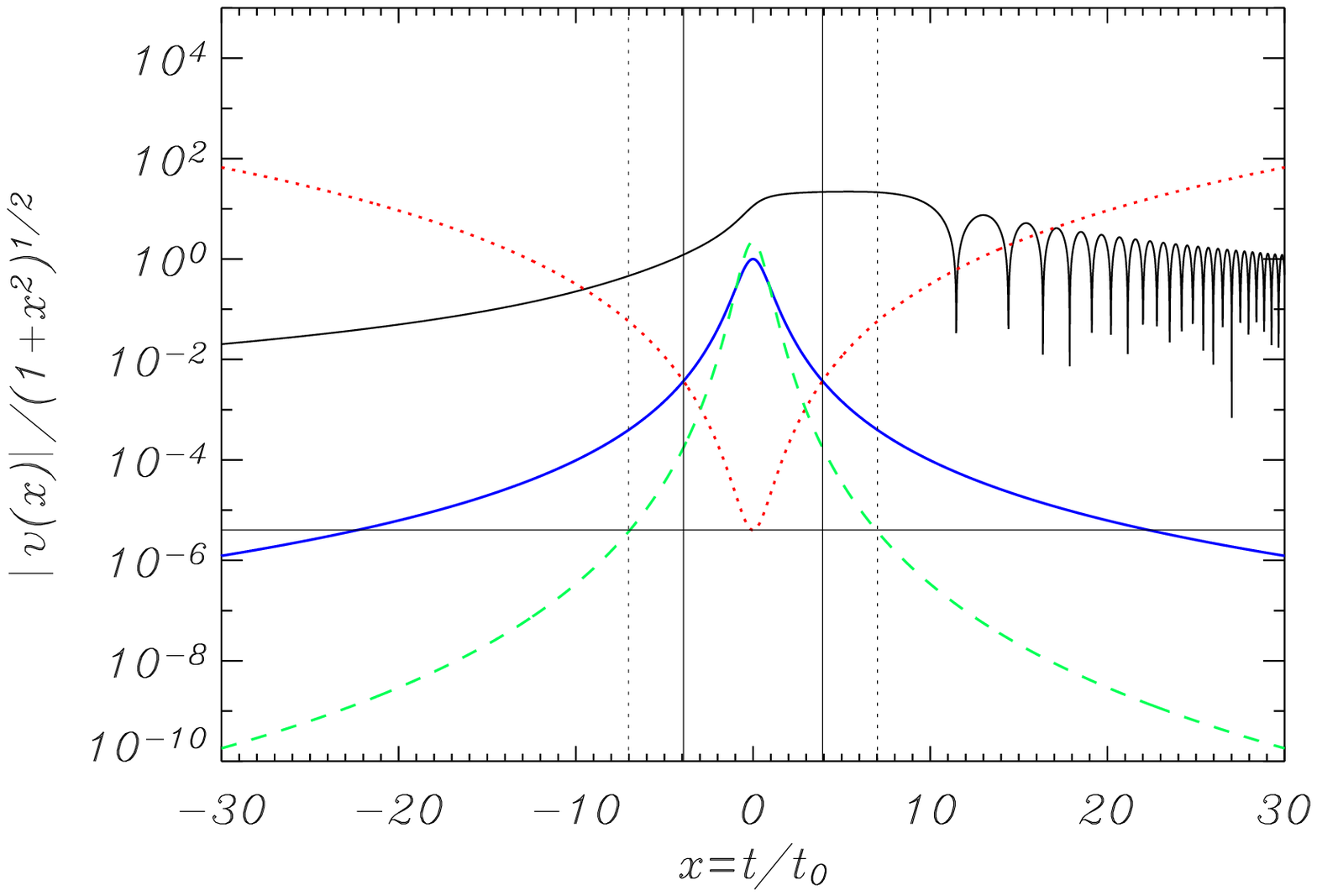}
\includegraphics[width=8.5cm]{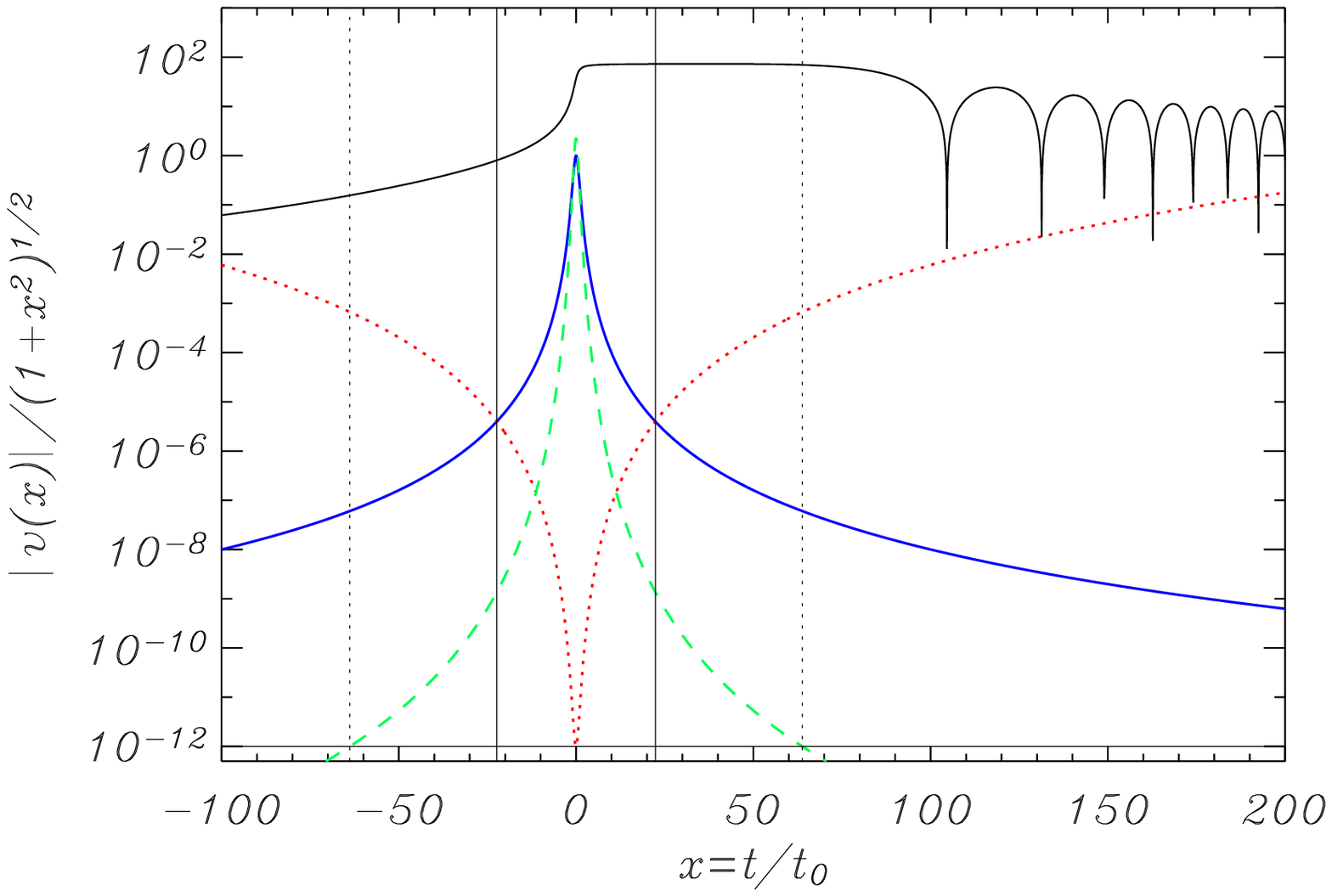}
\caption{Same as previous figures with $\omega=0.7$ and two different
wavenumbers $\tilde k = 2\times 10^{-3}$ (left) and $\tilde k =
10^{-6}$ (right).}
\label{fig:v07}
\end{figure*}

Our first task consists in going from the conformal time $\eta$ to the
more convenient time variable $t$ stemming from the change
(\ref{jauge}). With a dot indicating a derivative with respect to $t$,
the mode potential reads
\begin{equation}
\label{Vmu}
\frac{a''}{a} = a^{2(1-3\omega)} \left[\frac{\ddot a}{a} + \left(
  1-3\omega \right) \left( \frac{\dot a}{a} \right)^2 \right],
\end{equation}
and Eq.~(\ref{mu}) transforms into
\begin{widetext}
\begin{equation}
\label{v}
\label{nensecampeao}
\ddot v + \left[ \frac{k^2+2\mathcal{K}}{a^{2(1-3\omega)}} - \frac{3}{4} \left(
  1-3\omega \right) \left(1-\omega\right) \left( \frac{\dot a}{a}
  \right)^2 - \frac{3}{2}\left(1-\omega\right)\frac{\ddot a}{a} \right]
  v = 0,
\end{equation}
in which we have defined $v\equiv a^{\frac{1}{2}\left(
  1-3\omega\right)}\mu$. Specializing to the flat $\mathcal{K}=0$ case
  and setting
$$ v \equiv \lP\sqrt{t_0} \bar{v} \qquad \hbox{,} \qquad
  x\equiv\frac{t}{t_0} \qquad \hbox{and} \qquad \tilde k\equiv
  \frac{k}{\bar{k}_0} \qquad \hbox{with} \qquad \bar{k}_0 \equiv
  \left( t_0 a_0^{3\omega-1} \right)^{-1}, $$
\end{widetext}
we obtain
\begin{equation}
\label{vbar}
\frac{\dd^2\bar{v}}{\dd x^2} + \left[ \tilde k^2 \left( 1+x^2
\right)^{\frac{2\left(3\omega-1\right)}{3\left(1-\omega\right)}} -
\frac{1}{\left( 1+x^2 \right)^2} \right] \bar{v} = 0.
\end{equation}
which is in a useful form for the practical purpose of numerical
resolution. We shall assume the usual vacuum state initial condition for the
modes, \ie we set~\cite{MFB}
\begin{equation}
\label{mu_ini}
\mu_\mathrm{ini} = \frac{\sqrt{3} \lP}{\sqrt{k}} \exp \left[ -ik\left(
    \eta - \eta_\mathrm{ini}\right) \right],
\end{equation}
where $\eta_\mathrm{ini}$ is an arbitrary (and physically irrelevant
as was checked numerically) constant conformal time, which we set to
zero in what follows without loss of generality.
Figs.~\ref{fig:v01} to \ref{fig:v07} show the actual mode
calculated numerically with Eq.~(\ref{vbar}).

The power spectrum can now be defined as~\cite{MFB}
\begin{equation}
\label{PS}
k^3 \mathcal{P}_h \equiv \frac{2 k^3}{\pi^2} \left|
\frac{\mu}{a} \right|^2,\end{equation} leading to
\begin{equation}
\label{PSv}
k^3 \mathcal{P}_h = \frac{2 \tilde k^3}{\pi^2}\bar{k}_0^2
\frac{|\bar{v}|^2}{1+x^2} \left(\frac{\lP}{a_0}\right)^2,
\end{equation}
which, although in general being a time-dependent quantity, happens to
be constant in the expanding phase for the time period we are
interested in. Therefore, it suffices to solve Eq.~(\ref{vbar}) with
the initial condition (\ref{mu_ini}) to obtain the gravitational wave
power spectrum we are seeking. This is how we obtained the figures.

\subsection{Piecewise approximation and matching in the flat spatial
  section case}

\subsubsection{Asymptotic behaviors}

Given Eq.~(\ref{at}) and the relation (\ref{jauge}) between the two
time parameters, one has
\begin{equation}
\eta = a_0^{1-3\omega} \null_2\mathcal{F}_1 \left[\frac{1}{2},
\frac{3\omega-1}{3(\omega-1)}
,\frac{3}{2},-\left(\frac{t}{t_0}\right)^2 \right] t ,
\end{equation}
where $ \null_2\mathcal{F}_1 $ is a hypergeometric function and we
have assumed a common origin for both times (\ie $\eta=0$ for
$t=0$). This can be simplified by considering that one is mostly
interested, either for setting initial conditions or for observing the
resulting power spectrum, for times much larger than the typical
bounce duration, \ie for $t\gg t_0$. Then, one recovers the usual
perfect fluid power law solution for the scale factor, allowing us to
write
\begin{equation}
\label{keta}
k\eta = \pm \frac{3(1-\omega)}{1+3\omega} \tilde k
|x|^{\frac{1+3\omega}{3(1-\omega)}},
\end{equation}
where the sign is to be determined by that of $x$.

Eq.~(\ref{mu}) with $\Ka=0$ has a potential for gravitational waves
that is written $V_\mathrm{grav} = a''/a$, and which can be expressed
in terms of the $x$ variable as
\begin{equation}
\label{Vx}
V_\mathrm{grav} = A \left(1+x^2\right)^\frac{4}{3(\omega-1)} \left[
  \left(3\omega-1\right)x^2 +3\left(\omega-1\right)\right],
\end{equation}
where the constant $A$ is given by
$$ A = -\frac{2}{9}\left[\frac{a_0^{1-3\omega}}{(\omega-1)
t_0}\right]^2.$$ For large values of $\eta$, hence of $x$, and
provided $\omega\not=\frac{1}{3}$,
one gets
\begin{equation}
\label{Vasym}
V_\mathrm{grav} \sim (3\omega-1)Ax^\frac{2(3\omega+1)}{3(\omega-1)},
\end{equation}
which vanishes asymptotically for all cases of practical interest
($-1/3\leq \omega \leq 1$).

The case $\omega=1/3$ is a very special and simple one: the time $t$
is conformal time $\eta$, Eq.~(\ref{nensecampeao} is identical to
Eq.~(\ref{mu}), and $a(\eta) = a_0 [1+(\eta/\eta_0)^2]^{1/2}$. In a
different context, this same mode equation (with ${\mathcal{K}}=0$)
was treated in Ref.~\cite{jpps}, yielding an spectral index
$k^3\mathcal{P}_h \propto k^{n_{_\mathrm{T}}}$ given by
$n_{_\mathrm{T}} = 2$.

\subsubsection{Matching points}

Let us determine the end of potential domination point, denoted by
$x_{_\mathrm{M}}$ in what follows, i.e., the time at which $k^2 =
a''(x_{_\mathrm{M}})/a(x_{_\mathrm{M}})$. This is
\begin{equation}
\label{xend}
x_{_\mathrm{M}} = \left[ \frac{9\left(1-\omega\right)^2}{2|1-3\omega|}
\tilde k^2 \right]^{\frac{3 \left(\omega-1\right)}{2\left( 1+3\omega
\right)}}.
\end{equation}
This point will also be used to match different solutions and thus
propagate the mode through the bounce.

It is interesting to note that the corresponding point for the
evolution of $\bar{v}$, namely the point obtained by annihilating the
bracket in Eq.~(\ref{vbar}), called $x_\mathrm{exit}$, is
$$
x_\mathrm{exit} = \sqrt{\tilde{k}^{\frac{3}{2}(\omega-1)}-1}\simeq
\tilde{k}^{\frac{3}{4}(\omega-1)},
$$ so that the ratio is
\begin{equation}
\label{xoverx}
\frac{x_\mathrm{exit}}{x_{_\mathrm{M}}} \propto \tilde k^{\frac{9}{4}
  \frac{\left(1-\omega\right)^2}{1+3\omega}} \ll 1,
\end{equation}
which, for long wavelengths ($\tilde k \ll 1$) is much less than one
(recall that $-1/3 < \omega <1$). Therefore, the modes we consider
in the numerical solution have no time to start oscillating
before reaching $x_{_\mathrm{M}}$.

\subsubsection{Solutions}

Putting Eq.~(\ref{Vasym}) into Eq.~(\ref{mu}) and using (\ref{keta}),
we arrive at the conclusion that, sufficiently far from the bounce,
the perturbation mode satisfies
\begin{equation}
\label{Modes}
\mu'' +\left[ k^2 +\frac{2(3\omega-1)}{(1+3\omega)^2\eta^2}\right]\mu
= 0,
\end{equation}
whose solution is
\begin{equation}
\label{Bessel}
\mu = \sqrt{\eta} \left[ c_1(k) H^{(1)}_\nu (k\eta)+ c_2(k)
  H^{(2)}_\nu(k\eta)\right],
\end{equation}
with
$$ \nu = \frac{3(1-\omega)}{2(3\omega+1)}, $$ $c_1$ and $c_2$ being
two constants depending on the wavelength, $H^{(1,2)}$ being Hankel
functions.

This solution applies asymptotically, where one can impose initial
conditions on the mode, as well as in the matching region for which
$V_\mathrm{grav}\sim k^2$, provided $\tilde{k}^2\ll 1$. Demanding the
Bunch-Davies vacuum normalization (\ref{mu_ini}) then implies
$$ c_1=0 \quad \hbox{and} \quad c_2=\lP \sqrt{\frac{3\pi}{2}}
\ex^{-i\frac{\pi}{2} \left(\nu+\frac{1}{2}\right)}.
$$

The solution can also be expanded in powers of $k^2$ according to the
formal solution~\cite{MFB}
\begin{eqnarray}\label{solform}
\frac{\mu}{a}& - & \mathcal{O}(k^{j\geq 4}) = A_1(k)\biggl[1 - k^2
  \int^t \frac{\dd\bar \eta}{a^2\left(\bar \eta\right)}
  \int^{\bar{\eta}}
  a^2\left(\bar{\bar{\eta}}\right)\dd\bar{\bar{\eta}}\biggr]\nonumber
  \\ &+& A_2(k) \biggl[\int^\eta\frac{\dd\bar{\eta}}{a^2} - k^2
  \int^\eta \frac{\dd\bar{\eta}}{a^2} \int^{\bar{\eta}} a^2
  \dd\bar{\bar{\eta}} \int^{\bar{\bar{\eta}}}
  \frac{\dd\bar{\bar{\bar{\eta}}}}{a^2} \biggr],
\end{eqnarray}
where $A_1$ and $A_2$ are two constants depending only on the
wavenumber $k$ through the initial conditions. Neglecting the
$\mathcal{O}(k^2)$ terms, for the expanding phase, the $A_2$ term is
known as the decaying mode, and the power spectrum (\ref{PS}) can then
be approximated accurately by a constant; this constant power spectrum
is the one we are looking for. Although this form is particularly
valid as long as $k^2\ll a''/a$, \ie when the mode is below its
potential, Eq.~(\ref{solform}) should formally apply for all times.
In the matching region, the $\mathcal{O}(k^2)$ terms may give
contributions to the amplitude, but they do not alter the
$k$-dependence of the power spectrum.

For the solution (\ref{at}), the leading order of the solution
(\ref{solform}) reads
\begin{eqnarray}
\frac{\mu}{a} &=& A_1 + A_2 t_0 a_0^{3(\omega-1)} \arctan x \cr & \sim
& A_1-A_2 t_0 a_0^{3(\omega-1)} \left(
\frac{\pi}{2}+\frac{1}{x}\right),\cr \Longrightarrow \mu &\sim&\tilde
A_1 x^\frac{2}{3(1-\omega)} + \tilde A_2
x^\frac{3\omega-1}{3(1-\omega)},
\label{solmu0}
\end{eqnarray}
where in the last steps we have taken the limit $x\to-\infty$ and
identified the leading orders in $x$, with
$\tilde A_1 = A_1 a_0 -\frac{\pi}{2}a_0^{3\omega-2}t_0 A_2$ and
$\tilde A_2=-a_0^{3\omega-2}t_0 A_2$. Propagating this solution on the
other side of the bounce, in the expanding epoch, yields the required
power spectrum, \ie the limit for $x\to +\infty$, namely
\begin{equation}
\label{specEND}
\frac{\mu}{a}\bigg|_\mathrm{const} \sim A_1 + \frac{\pi}{2}
  a_0^{3(\omega-1)} t_0 A_2 = \frac{1}{a_0} \left( \tilde A_1 - \pi
  \tilde A_2\right),
\end{equation}
where we have taken only the constant part of the modes.

\subsubsection{Matching and spectrum}

In order to get the spectrum of gravitational waves produced by our
bouncing model, it suffices to match $\mu$ and $\mu'$ at
$\eta_{_\mathrm{M}}$, corresponding to $x_{_\mathrm{M}}$ given by
(\ref{xend}) through (\ref{keta}). At this point, the mode function
(\ref{Bessel}) and its derivative read
\begin{equation}\label{mum}
\mu(\eta_{_\mathrm{M}}) = \frac{C}{\sqrt{k}} \quad \hbox{and} \quad
  \mu'(\eta_{_\mathrm{M}}) = D\sqrt{k},
\end{equation}
where the constants $C$ and $D$ are given by
\begin{equation}
C=c_2\sqrt{k\eta_{_\mathrm{M}}} H_\nu^{(2)}(k\eta_{_\mathrm{M}}),
\end{equation}
and
\begin{eqnarray} D &=& \frac{c_2}{2} \Bigg\{
\frac{H_\nu^{(2)}(k\eta_{_\mathrm{M}})}{\sqrt{k\eta_{_\mathrm{M}}}}
\cr & & \hskip3mm +
\sqrt{k\eta_{_\mathrm{M}}} \left[ H_{\nu-1}^{(2)}(k\eta_{_\mathrm{M}})
- H_{\nu+1}^{(2)}(k\eta_{_\mathrm{M}})\right] \bigg\},
\end{eqnarray}
with
$$k\eta_{_\mathrm{M}}=\frac{\sqrt{2|1-3\omega|}}{1+3\omega}.$$ This is
also expressed as
\begin{equation}
\mu (\eta_{_\mathrm{M}}) = \frac{\tilde C}{\sqrt{\tilde k}} \quad
  \hbox{and} \quad \mu'(\eta_{_\mathrm{M}}) = \tilde D\sqrt{\tilde k},
\end{equation}
with $\tilde C = \sqrt{t_0} a_0^{(3\omega-1)/2} C$ and $\tilde D =
  a_0^{(1-3\omega)/2} D/\sqrt{t_0}$.

\begin{figure*}[t]
\includegraphics[width=8.5cm]{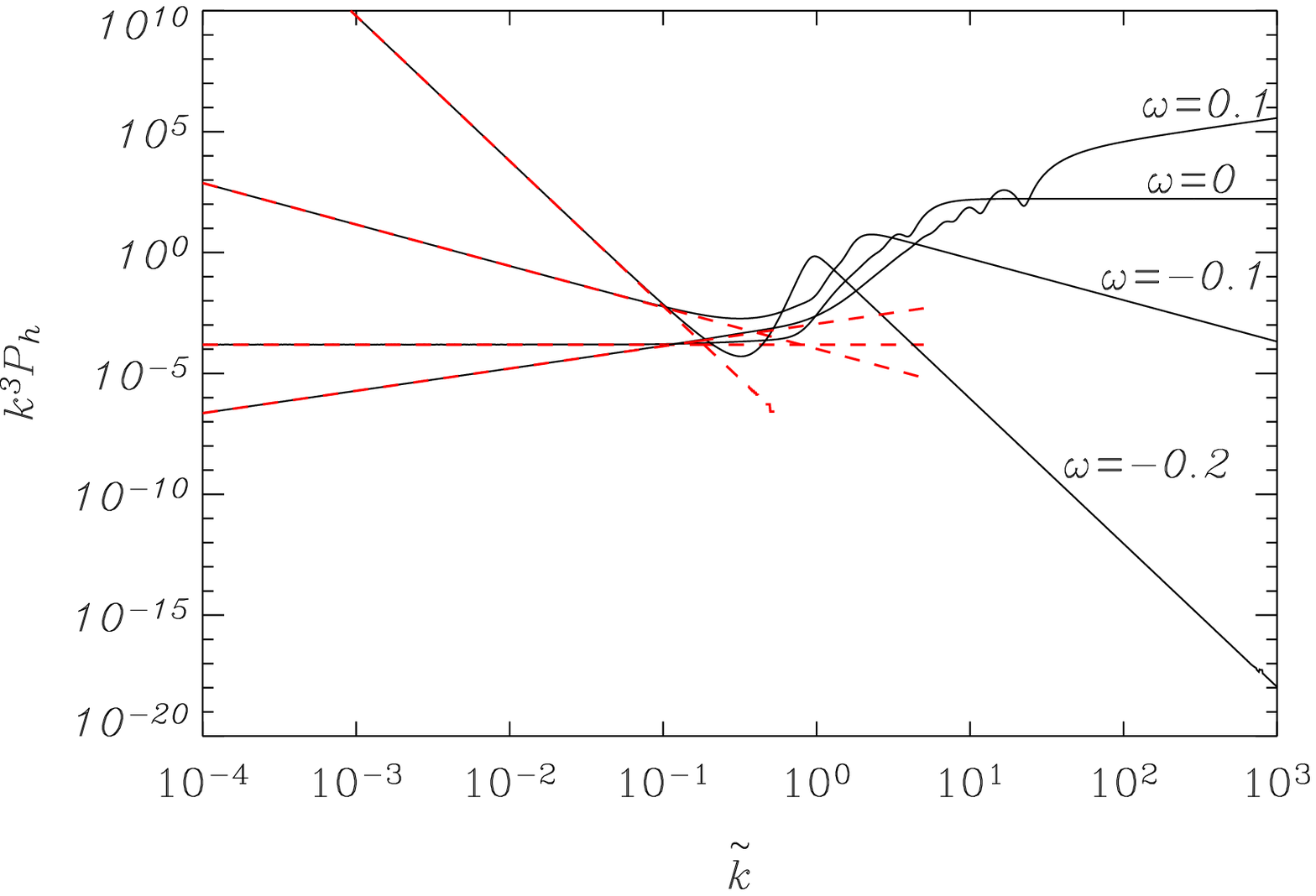}
\includegraphics[width=8.5cm]{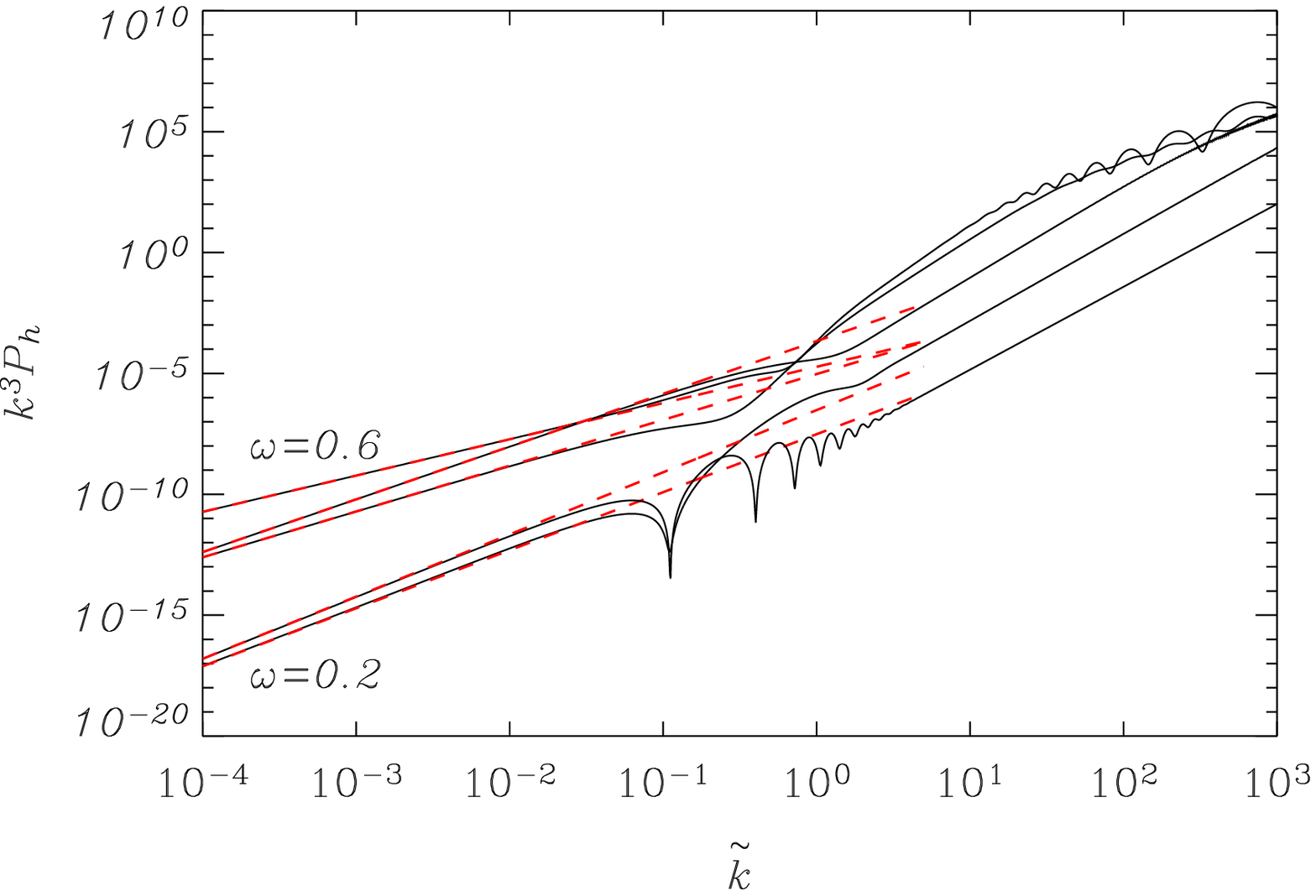}
\caption{Power spectra for different values of the state parameter
$\omega$. The dashed lines represent the approximation (\ref{powspec})
with the power index given by Eq.~(\ref{index}), while the full lines
are the spectra obtained by numerically solving
Eq.~(\ref{vbar}). Shown on the right panel are solutions for
$\omega\in \{ 0.2, 0.3, 0.4, 0.5, 0.6\}$.}
\label{fig:spectrum}
\end{figure*}

Matching $\mu$ [Eqs.~(\ref{solmu0}) and (\ref{mum})] and its
derivative with respect to conformal time, namely $\mu'=a^{1-3\omega}
t_0^{-1} \dd\mu/\dd x$, one obtains, to leading order
\begin{eqnarray}
\tilde A_1 &=& \left[ \frac{3\omega-1}{3\alpha\left(\omega-1\right)}
  \tilde C + a_0^{3\omega-1} t_0 \beta \tilde D \right] \tilde
  k^\frac{3\left(1-\omega\right)}{2\left(3\omega+1\right)},\label{A1}\\
  \tilde A_2 &=& \left[ \frac{2}{3\beta\left(1-\omega\right)} \tilde C
  - a_0^{3\omega-1} t_0 \alpha \tilde D \right] \tilde
  k^\frac{3\left(\omega-1\right)}{2\left(3\omega+1\right)},\label{A2}
\end{eqnarray}
with
$$ \alpha = \left[
\frac{9\left(1-\omega\right)^2}{2\left|1-3\omega\right|}
\right]^{-1/(1+3\omega)}
$$
and 
$$ \beta = \left[
\frac{9\left(1-\omega\right)^2}{2\left|1-3\omega\right|}
\right]^{\frac{1-3\omega}{2(1+3\omega)}}.
$$

The coefficients $\tilde A_1$ and $\tilde A_2$ contain each a
power-law behaviors in $k$. Because $\omega<1$, the power in $\tilde
A_2$ [Eq.~(\ref{A2})] is negative definite and that in $\tilde A_1$
[Eq.~(\ref{A1})] is positive definite. Therefore, $\tilde A_2$ is the
dominant mode and gives the spectral index, while $\tilde A_1$
provides the sub-dominant mode that happens, incidentally, to
correspond to an unaltered propagation of the initial conditions. One
then gets the spectrum (\ref{specEND}), and finally the spectral index
$n_{_\mathrm{T}}$ writing
\begin{equation}
k^3\mathcal{P}_h \propto k^{n_{_\mathrm{T}}},
\label{powspec}
\end{equation}
and we get
\begin{equation}
\label{index}
n_{_\mathrm{T}} = \frac{12\omega}{1+3\omega}
\end{equation}
Note that the limit $\omega\rightarrow \frac{1}{3}$ of
Eq.~(\ref{index}) gives the correct index for radiation (see
Ref.~\cite{jpps}), although the calculation, in this case, should not
be valid; this is due to the expected continuity of the spectral index
with the equation of state.  The spectrum as calculated numerically is
plotted on Fig~\ref{fig:spectrum} for various values of $\omega$
together with the approximation (\ref{index}).

It is interesting to note that this result was also presented in the
semiclassical approximation (classical background and quantum
perturbations) in Ref.~\cite{sta}. In Ref.~\cite{sta}, the asymptotic
behaviors both in the past and future infinities are two, possibly
different, power laws for the contraction and expansion phases,
whereas the type of bounces we studied here is restricted to equal
asymptotic behaviors, i.e., for $|t|\gg t_0$. Since the potentials in
the equations for $\mu$ are smooth and large compared to $k$ around
the bounce, it looks like the full quantum effects and details of the
bounce do not change significantly the main spectral features of the
gravitational wave produced. It would be interesting to verify if this
results still holds for other bounces, e.g. those having different
asymptotic behaviors and/or more complicated shapes of the potential
for $\mu$. In this last situation, and if the results of
Ref.~\cite{jmpp} apply, one would expect the actual spectra to be
different.

\subsection{Power spectrum for curved spatial sections and radiation}

In this subsection we consider the power spectrum of tensor
perturbations for quantum cosmological backgrounds with curved
spatial sections. As mentioned in Sec.~III, only in the radiation
case one can obtain analytic solutions for the quantum background.
Hence, we will restrict ourselves to this fluid from now on.

Inserting Eq.~(\ref{atk}) into Eq.~(\ref{mu}), and noting that
$k^2=m^2-3\mathcal{K}$, where $m$ is an integer greater or equal to $3$
for $\mathcal{K}=1$, and a real number greater than zero for
$\mathcal{K}=-1$, we obtain

\begin{equation}
\label{muk}
\mu''+\left\{m^2 -\frac{\eta_0^2}
{[\eta_0^2\mathcal{K}\cos^2(\sqrt{\mathcal{K}}\eta)+
\sin^2(\sqrt{\mathcal{K}}\eta)]^2} \right\}\mu =0.
\end{equation}
The effective potential 
\begin{equation}
\label{VeffK}
V_\mathrm{eff} = \frac{\eta_0^2}
{[\eta_0^2\mathcal{K}\cos^2(\sqrt{\mathcal{K}}\eta)+
\sin^2(\sqrt{\mathcal{K}}\eta)]^2}
\end{equation}
has one maximum given by $1/\eta_0^2$ and goes to zero when
$\eta\rightarrow\infty$ for $\mathcal{K}=-1$. It oscillates between
$1/\eta_0^2$ and $\eta_0^2$, which are respectively a maximum and a
minimum provided $\eta_0 < 1$, when $\mathcal{K}=+1$. Indeed, as we
have seen in Sec.~\ref{sec:traj}, in order for the background models
to be realistic, one must have $\eta_0 \ll 1$. Hence, the maxima of
the effective potential are very high in both cases and the minima are
very small in the $\mathcal{K}=1$ case.  Large wavelengths (small $m$)
will cross the effective potential and the perturbations will be
amplified at each bounce. This induces an instability of the model
because this enhancement happens an infinite number of times, and
therefore, however small the initial perturbation might have been,
there is a time at which the linear theory is no longer valid and the
cosmological setup breaks down.

\begin{figure}[t]
\includegraphics[width=8.5cm]{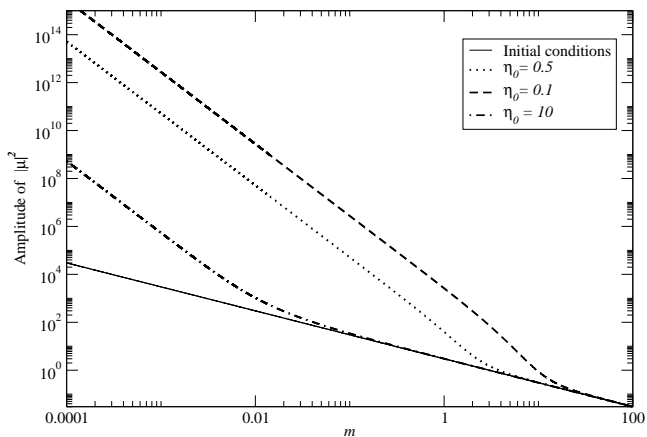}
\caption{Amplitude amplification of the gravitational modes in the
  $\mathcal{K}=-1$ curved case filled with radiation. For $m\ll 1$,
  the amplitude behaves, as expected at least for the case $\eta_0\ll
  1$, as a power law since then $\mu\propto m^{-3/2} \sin{m\eta}$.}
\label{fig:muKm1}
\end{figure}

For $\mathcal{K}=-1$, the situation is very similar to the flat
case. The conformal time $(\eta_{_\mathrm{M}}$ of potential crossing
is given as the solution of the equation
\begin{equation}
\label{etak0}
\sinh(\eta_{_\mathrm{M}})=\pm\sqrt{\biggl(\frac{1}{m\eta_0}-1\biggr)}
\frac{\eta_0}{\sqrt{1+\eta_0^2}},
\end{equation}
which has a real solution provided $m\eta_0<1$. Thus,
the mode crosses the potential only in this case. This constraint is
however satisfied for the situations we are interested in, namely,
$\eta_0\ll 1$ and $m\ll 1$. One then obtains
\begin{equation}
\label{etak}
|\eta_{_\mathrm{M}}|\approx\mathrm{arcsinh}
\biggl(\sqrt{\frac{\eta_0}{m}}\biggr).
\end{equation}
We have two limiting cases, namely $\eta_0\gg m$ and $\eta_0\ll m$
yielding, respectively,
\begin{equation}
\label{etaMg}
|\eta_{_\mathrm{M}}|\approx \ln[2(\eta_0/m)^{1/2}],
\end{equation}
and
\begin{equation}
\label{etaMp}
|\eta_{_\mathrm{M}}|\approx (\eta_0/m)^{1/2}.
\end{equation}

The effect of the potential for $\mu$ is to increase its amplitude by
a factor shown on Fig.~\ref{fig:muKm1}, as well as to mix the
exponential terms. This can be easily seen by the following
approximation. For $\eta_0\ll 1$, the maximum of the effective
potential at $\eta=0$ is very large while for $\eta\gg\eta_0$ it
behaves like $\eta_0^2/\sinh ^4(\eta)$, which goes to zero when
$\eta_0\rightarrow 0$.  Hence the effective potential (\ref{VeffK})
can be well approximated to a Dirac delta in this limit. Its
integration reads
\begin{eqnarray}
\int V_\mathrm{eff}(\eta)\dd\eta &=& \frac{1}{\eta_0^2} \int\dd\eta
\left[ 1+\left( 1+\frac{1}{\eta_0^2} \right) \sinh^2 \eta
\right]^{-2}\cr & & \cr & =& \frac{\left(\eta_0^2+1\right)
\sinh\left(2\eta\right)}{2\left[\left(\eta_0^2-1\right) +
\left(\eta_0^2+1\right) \cosh\left(2\eta\right)\right]} \cr &&\cr & &
+ \frac{1-\eta_0^2}{2\eta_0} \tan^{-1}
\left(\frac{\tanh\eta}{\eta_0}\right),
\label{VeffInt}
\end{eqnarray}
which implies
\begin{equation}
\int_{-\infty}^{\infty} V_\mathrm{eff} (\eta) \dd\eta = 1+
\frac{1-\eta_0^2}{\eta_0} \tan^{-1}\frac{1}{\eta_0} \simeq
\frac{\pi}{2\eta_0} + \mathcal{O}\left(\eta_0\right),
\end{equation}
where in the last step we have assumed that $\eta_0\ll 1$.

Hence, one can approximate the effective potential by a Dirac
distribution~\cite{jmpp} $\pi\delta (\eta)/2\eta_0$. The solution
for $\eta\not=0$ is then simply Eq.~(\ref{mu_ini}) with $k$
substituted by $m$ for $\eta < 0$, and $\mu =
A\ex^{im\eta}+B\ex^{-im\eta}$ for $\eta>0$. Demanding that $\mu$ be
continuous across the potential and imposing Eq.~(\ref{muk}) then
leads to another matching condition, namely
\begin{equation}
\mu'(0^+) - \mu'(0^-) = \frac{\pi}{2\eta_0} \mu(0).
\end{equation}
One then finds that
\begin{eqnarray}
A &=& - \frac{i\pi\sqrt{3} \lP}{4\eta_0 m^{3/2}}, \nonumber\\
B &=& \frac{i\pi\sqrt{3} \lP}{4\eta_0 m^{3/2}}+\sqrt{\frac{3}{m}}\lP,
\label{coef}
\end{eqnarray}
and finally, in the long wavelength approximation for which $m\ll 1$,
that $\mu \propto m^{-3/2} \sin{m\eta}$.  This is exactly what is
obtained numerically, as shown on Fig.~\ref{fig:muKm1}. Note also that
when the curves reach the $m\eta_0 >1$ region, the amplitude is the
initial one: the mode has not crossed the potential as explained
above.

Note incidentally that solution (\ref{coef}) is exatly the same as the
one obtained in Ref.~\cite{jpps} and in the present paper for
radiation with ${\mathcal K}=0$, where we used the matching
method. This is because also in this case one can approximate the
potential to a Dirac distribution as $V\approx \eta_0^2/\eta^4$ when
$\eta\gg\eta_0$, which goes to zero in the limit $\eta_0\rightarrow
0$. However, for the parabolic scale factor
$a(\eta)=a_0[1+(\eta/\eta_0)^2]$ also treated in Ref.~\cite{jpps},
which solutions are quite different from (\ref{coef}), the potential
is $V=1/(\eta^2+\eta_0^2)$, whose limit for $\eta\gg\eta_0$ is
$1/\eta^2$, independent of $\eta_0$. Hence in this case, the effective
potential cannot be approximated by a Dirac distribution, and the
final spectrum is very different. We thus confirm that the power
spectrum of perturbations through a bounce may depend significantly of
the details of the bounce itself.

With the coefficients (\ref{coef}), one can calculate the spectrum
\begin{equation}
\label{PS2}
m^3 \mathcal{P}_h \equiv \frac{2 m^3}{\pi^2} \left|
\frac{\mu}{a} \right|^2,\end{equation}
for the two possible matching points (\ref{etaMg},\ref{etaMp}),
yielding  $m^3 \mathcal{P}_h\propto m^3\ln^2 (m)$ and
$m^3 \mathcal{P}_h=m^2$. Note that, as expected, the case
$\eta_0\ll m$ yields the same spectrum as the flat case: the
two scale factors are quite similar in that limit.

\section{Conclusion}

We have obtained the power spectrum of tensor perturbations in
bouncing quantum cosmological models with a perfect fluid satisfying
$p=\omega\rho$ for flat spatial sections and $-\frac{1}{3} < \omega <
1$, and for curved spatial sections with $\omega =\frac{1}{3}$. For
flat spatial sections, the spectral index for large wavelengths is
$n_{_\mathrm{T}} = 12\omega/(1+3\omega)$.  The positive curved spatial
section model is unstable, while the negative curved spatial section
model amplifies the modes, changing the amplitude to a power index of
$n_{_\mathrm{T}} \approx 3 $ or $n_{_\mathrm{T}}=2$, depending on the
parameters.  All cases lead to oscillations in the primordial
spectrum.

The most interesting case is the one of radiation, which is the best
perfect fluid model for the early Universe (all particles are
ultra-relativistic).  For almost flat spatial sections we have
$n_{_\mathrm{T}}\approx 2$, which is different from the predictions of
inflation. Hence, this model can be potentially tested against
inflation in future observations, specially concerning the
polarization of the cosmic microwave background, Planck observations,
and gravitational wave detectors if we calculate the amplitude of
these perturbations.

The next step would be to calculate the spectrum of scalar
perturbations of these models. The dynamical equations for scalar
perturbations are not, however, as simple as Eq.~(\ref{mu}). The steps
we have taken in Sec.~(2) in order to arrive at Eq.~(\ref{mu}) in the
case where the background is also quantized are not so simple in the
case of scalar perturbations, specially due to the matter terms. This
is work in progress. Attainment of the power spectrum of scalar
perturbations is crucial not only to test the model against WMAP
observations, but also to calibrate and obtain the precise spectrum of
tensor perturbations for possible comparisons with LIGO and VIRGO
future data.

\section{Acknowledgments}

We would like to thank CNPq of Brazil for financial support. We would
also like to thank both the Institut d'Astrophysique de Paris and the
Centro Brasileiro de Pesquisas F\'{\i}sicas, where this work was done,
for warm hospitality. We very gratefully acknowledge various
enlightening conversations with J\'er\^ome Martin. We also would like
to thank CAPES (Brazil) and COFECUB (France) for partial financial
support. 
%We finally acknowledge a referee for pointing us Ref.~\cite{sta}.

\end{document}